\documentclass[aps,prl,reprint,superscriptaddress,longbibliography]{revtex4-2}
\newcommand{\BM}{\boldsymbol}

\newcommand{\hz}[1]{{\color{violet}{#1}}}

\usepackage{amsthm}
\usepackage{amssymb}
\usepackage{amsmath}
\usepackage{graphicx}
\usepackage{setspace}
\usepackage{enumerate}
\usepackage{multirow}
\usepackage{ragged2e}
\usepackage{xcolor}

\usepackage{hyperref}% add hypertext capabilities
\graphicspath{{Letter_Picture/}}%指定图片文件夹路径(Path to the picture folder)

\begin{document}

\preprint{APS/123-QED}

\title[Manuscript]{Nonequilibrium hysteretic phase transitions in periodically light-driven superconductors}
% Force line breaks with \\

\author{Huanyu Zhang}
\email{zhang@dyn.phys.s.u-tokyo.ac.jp}
\affiliation{Department of Physics, The University of Tokyo, Hongo, Tokyo 113-0033, Japan}
\author{Kazuaki Takasan}
\email{kazuaki.takasan@phys.s.u-tokyo.ac.jp}
\affiliation{Department of Physics, The University of Tokyo, Hongo, Tokyo 113-0033, Japan}
\author{Naoto Tsuji}
\email{tsuji@phys.s.u-tokyo.ac.jp}
\affiliation{Department of Physics, The University of Tokyo, Hongo, Tokyo 113-0033, Japan}
\affiliation{RIKEN Center for Emergent Matter Science (CEMS), Wako 351-0198, Japan}

\date{\today}% It is always \today, today,
             %  but any date may be explicitly specified

\begin{abstract}
    We find nonequilibrium phase transitions accompanied by multiple (nested) hysteresis behaviors in superconductors coupled to baths under a time-periodic light driving.
    The transitions are demonstrated with a full phase diagram in the domain of the driving amplitude and frequency by means of the Floquet many-body theory. 
    In the weak driving regime with a frequency smaller than half of the superconducting gap, excited quasiparticles are accumulated at the far edges of the bands,
    realizing a distribution reminiscent of the Eliashberg effect, 
    which suddenly becomes unstable in the strong driving regime due to multi-photon-assisted tunneling across the gap mediated by the in-gap Floquet sidebands.
    We also show that superconductivity is enhanced in the weak driving regime without effective 
    cooling, which is attributed to the modulation of the spectrum due to Floquet sidebands.
\end{abstract}

%\keywords{Suggested keywords}%Use showkeys class option if keyword
                              %display desired
\maketitle
%\tableofcontents

\textit{Introduction.}---
Driving quantum phases of matter with light has been a promising route to control electronic states of materials in a nonthermal way \cite{aoki_nonequilibrium_2014, oka_floquet_2019, delaTorre2021, Murakami2023}, 
which may provide various possibilities of phase transitions that are otherwise elusive to access within thermal equilibrium. 
This is of particular interest in the context of superconductivity, since thermal phase transitions to superconducting states have been limited, to date, to low temperatures at ambient pressure. 
In fact, recent progress of intense laser techniques has allowed one to monitor real-time dynamics of superconducting states in a fast timescale (see, e.g., \cite{Giannetti2016, shimano_higgs_2020, Demser2020}) 
and observe light-induced superconducting-like states \cite{fausti_light-induced_2011, kaiser_optically_2014, Hu2014, mitrano_possible_2016, Niwa2019, Zhang2020, Buzzi2020, budden_evidence_2021, Isoyama2021, Katsumi2023, Fava2024, Nishida2024}, 
which may provide an opportunity of finding novel nonequilibrium phase transitions in light-driven superconductors.

During irradiation of multi-cycle pump light, the system is expected to approach a Floquet state \cite{Shirley1965, Sambe1973, Tsuji2024} (i.e., a time-periodic steady state) that may accommodate external-field-dressed quasiparticles,
which have been observed as Floquet-Bloch bands in time-resolved angle-resolved photoemission spectroscopy (tr-ARPES) experiments \cite{Wang2013, mahmood_selective_2016, Ito2023, Zhou2023, Merboldt2024, Choi2024}.
Signatures of Floquet states have also been glimpsed in transport \cite{McIver2020, Liu2025}, transient absorption spectroscopy \cite{Sie2015, Kobayashi2023}, nonlinear optical response \cite{Shan2021}, and tunneling spectroscopy \cite{park_steady_2022, haxell_microwave-induced_2023} measurements.
When the frequency $\Omega$ of the driving field is much larger than any other relevant energy scales, the system is well described by an effectively renormalized Hamiltonian, leading to the Floquet engineering \cite{bukov_universal_2015, oka_floquet_2019, Weitenberg2021}. This effect can be used, for example, to enhance superconductivity by suppressing hopping amplitudes 
\cite{ido_correlation-induced_2017, Dasari2018}. On the other hand, it is more complicated to understand the behavior of superconducting states at low-frequency regimes.

\begin{figure}[htbp]%htbp
    \centering
    \includegraphics[width=0.99\linewidth]{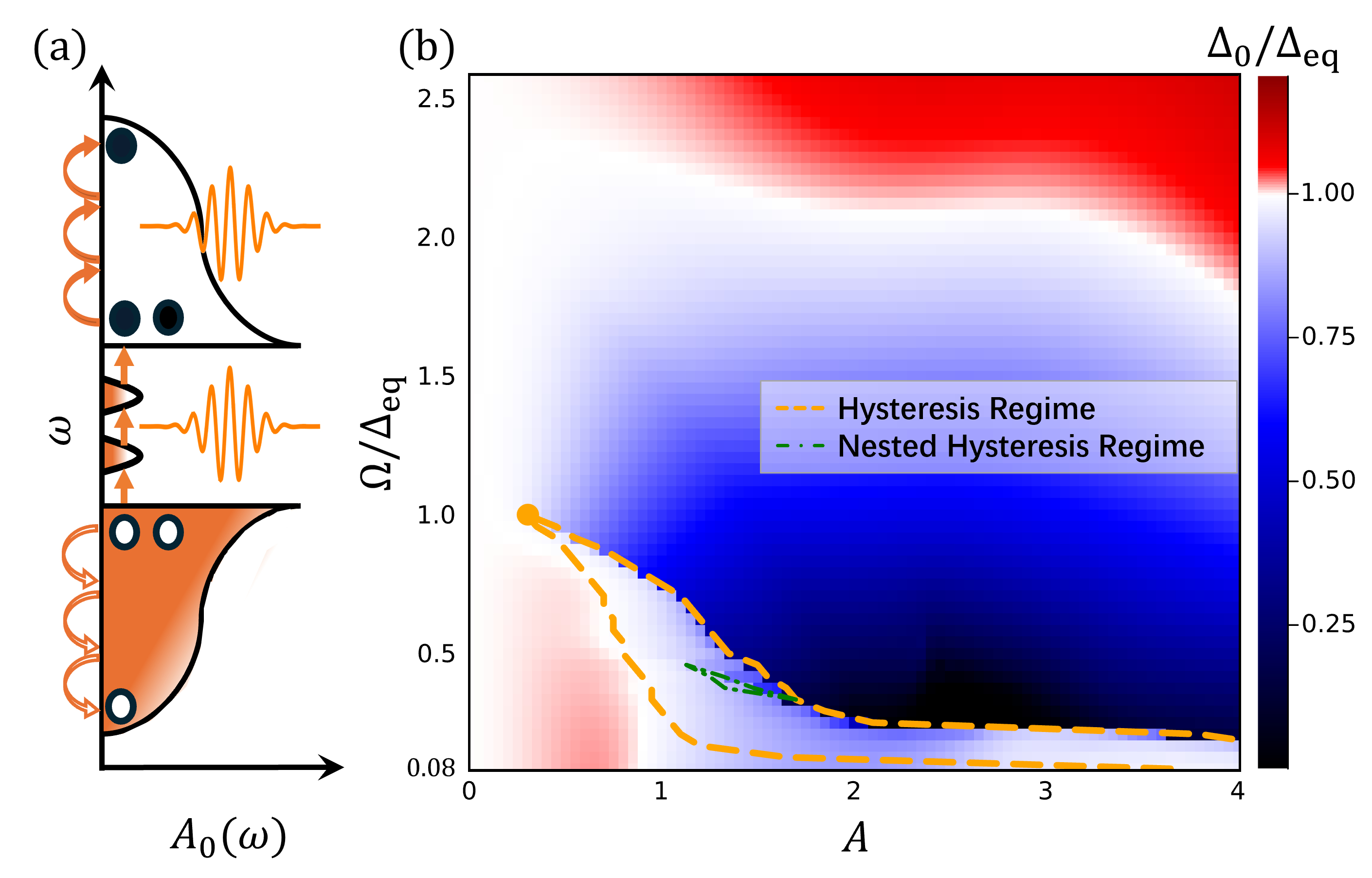}
    \caption{
    (a) Schematic picture of quasiparticles (holes) in superconductors driven by light shown by filled (empty) circles. Intraband scattering leads to an effective cooling of the system (Eliashberg effect), while there can occur multi-photon-assisted tunneling across the gap through Floquet sidebands (small peaks inside the gap).
    (b) Phase diagram of a BCS superconductor coupled to baths under a time-periodic light driving in the domain of the driving amplitude $A$ and frequency $\Omega$. 
    The color code plots the time-averaged superconducting order parameter $\Delta_0$ rescaled by the equilibrium one $\Delta_{\rm eq}$. 
    The area surrounded by the orange (green) curve indicates the (nested) hysteresis regime.
    The phase diagram is computed with $U=6.0$, $\Gamma=0.1$, and $T=0.2$, which gives $\Delta_{\rm eq}\approx 2.4$.}
    \label{Fig.1}
\end{figure}

When $\Omega$ becomes much smaller than the superconducting gap $2\Delta$,
thermally excited quasiparticles/holes are kicked away from the gap region \hz{(see Fig.~\ref{Fig.1}(a))}, resulting in an effective cooling of the quasiparticle distribution that may enhance superconductivity. This is known as the Eliahsberg effect \cite{eliashberg_film_1970,ivlev_influence_1971, Demser2020,ivlev_nonequilibrium_1973}, which has been observed in early microwave driving experiments \cite{wyatt_microwave-enhanced_1966,dayem_behavior_1967}.
While this mechanism is plausible, the effect of the modulation of the spectral function, which is beyond the quasithermal description \cite{delaTorre2021} 
but will be important for large driving amplitudes with sub-gap frequencies
%at the presence of large driving amplitudes with sub-gap frequencies
(not too far smaller than $2\Delta$), has not been considered before. Furthermore, the realized steady-state distribution of quasiparticles might be unstable: The kicked quasiparticles (holes) are accumulated at the far edge of the upper (lower) band, which causes significant population inversion (Fig.~\ref{Fig.1}(a)). 
It has not been clear how the quasiparticle distribution collapses if one increases the driving amplitude beyond the perturbative regime, and whether there exists a nonequilibrium phase transition in between the weak- and strong-driving regimes.

In this Letter, we revisit the problem of light-driven superconductors at sub-gap frequencies, and address the issue of the ultimate fate of the Eliashberg-type distribution of quasiparticles.
To this end, we solve nonequilibrium steady states (NESS) for a model of light-driven BCS superconductors coupled to baths by means of the Floquet Green's function method \cite{Faisal1989,Martinez2003,tsuji_correlated_2008,tsuji_nonequilibrium_2009,aoki_nonequilibrium_2014}, which can take account of both the effects of nonequilibrium distributions and modulated band structures.
The obtained phase diagram is shown in Fig.~\ref{Fig.1}(b).
We find multiple (nested) hysteretic phase transitions between the weak and strong driving regimes, which indicate characteristic features of the instability of the Eliashberg-type distribution that collapses due to multi-photon-assisted tunneling through in-gap Floquet sidebands (Fig.~\ref{Fig.1}(a)). 
Moreover, in the weak driving regime we show that superconductivity can be enhanced without effective cooling, 
which is attributed to the light-induced modulation of the spectrum due to Floquet sidebands
(for other mechanisms of light-induced enhancement of superconductivity, see, e.g., Refs.~\cite{Tsuji2011, Denny2015, Sentef2016, Okamoto2016, Komnik2016, Kennes2017, Sentef2017, Babadi2017, murakami_nonequilibrium_2017, Coulthard2017, Mazza2017, ido_correlation-induced_2017, Nava2018, Werner2018, Fabrizio2018, Dasari2018, Kaneko2019, Bittner2019, Werner2019b, Werner2019, Ray2023, Eckhardt2024, Ray2024,wang_universal_2024}).

\textit{Model and Formalism.}---
We consider a BCS superconductor on a square lattice illuminated by external continuous-wave light,
% (Fig.\ref{Bath1}(a))
whose Hamiltonian is given by
\begin{equation}
\begin{aligned}
  H_{\rm{BCS}}(t)=\sum_{\BM{k},\sigma} & \varepsilon_{\BM{k}-\frac{e}{\hbar}\BM{A}(t)} c_{\BM{k}\sigma}^\dagger c_{\BM{k}\sigma}% \\
   -\frac{U}{N} \sum_{\BM{k},\BM{p}} c_{\BM{k}\uparrow}^\dagger c_{-\BM{k}\downarrow}^\dagger  c_{-\BM{p}\downarrow} c_{\BM{p}\uparrow},
   \end{aligned}
\end{equation}
where $c_{\BM{k}\sigma}$ and $c^{\dagger}_{\BM{k}\sigma}$ stand for the electron annihilation and creation operators with momentum $\BM{k}$ and spin $\sigma \in \{\uparrow,\downarrow\}$, 
$U (>0)$ is the pairing strength between electrons, and $N$ is the number of lattice sites.
The band dispersion is given by $\varepsilon_{\BM{k}}=-2t_{\rm hop}\left[\cos(k_1a)+\cos(k_2a)\right]$, 
in which the vector potential $\BM{A}(t)=A\cos(\Omega t)\BM{e}_p$ is introduced via the Peierls substitution ($\BM{k}\rightarrow \BM{k}-\frac{e}{\hbar}\BM{A}(t)$, $e$ is the electric charge) \cite{peierls1933theorie}. 
The system is set at half filling with the chemical potential $\mu=0$.
We take the polarization along the diagonal direction, $\BM{e}_{p}=\frac{1}{\sqrt{2}}(1,1)$, 
and $A$ and $\Omega$ represent the driving amplitude and frequency, respectively.
To study the superconducting order, we define the gap function
%introduce the BCS-type uniform mean field
$\Delta(t) = - \frac{U}{N}\sum_{\BM k} \langle c_{-\BM{k} \downarrow}(t) c_{\BM{k} \uparrow}(t) \rangle$ to decouple the interaction term.
We use the units with $\hbar=1, e=1, t_{\rm hop}=1$, and the lattice constant $a=1$.

We assume that the superconductor is coupled to heat baths, and that the system arrives at a time-periodic steady state in the long-time limit where energy injected from the external driving field is balanced with dissipation. 
For a model of heat baths, we employ the free-fermion bath \cite{buttiker_small_1985,tsuji_nonequilibrium_2009} with temperature $T$ and the system-bath coupling $\Gamma$. 
The bath's degrees of freedom can be analytically integrated out, and the resulting effect of the bath can be incorporated in the self-energy, 
whose retarded component is given by $\Sigma_{\BM{k}}^{R}(\omega)=-i\Gamma $, 
and the lesser component is given by the fluctuation-dissipation theorem, $\Sigma_{\BM{k}}^{<}(\omega)=-2if^{\rm FD}_{T}(\omega){\rm Im}\,\Sigma_{\BM{k}}^{R}(\omega)$.
Here, $f^{\rm FD}_{T}(\omega)$ is the Fermi-Dirac distribution at temperature $T$.
In this setup, there is only an energy exchange without net particle current between the baths and the superconductor (i.e., nonequilibrium superconductivity in an energy mode \cite{Demser2020,tinkham2004introduction}).
This is to be distinguished from the previous study \cite{yang_intrinsic_2021} where superconductors are driven by the chemical potential change (i.e., charge mode \cite{Demser2020,tinkham2004introduction}).

The single-particle Green's function is then derived from the Dyson-Keldysh equation 
 \cite{jauho_time-dependent_1994,haug2008quantum,Eckstein2018} extended to the Nambu space,
\begin{gather}
    \label{Eq.Dyson1}
       \left( G_{\BM{k}}^{R}\right)^{-1}=\left( G_{0\BM{k}}^{R}\right)^{-1}-\Sigma_{\BM{k}}^{R},\\
    \label{Eq.Kel1}
    G^{<}_{\BM{k}}=G^{R}_{\BM{k}}*\Sigma^{<}_{\BM{k}}*G^{A}_{\BM{k}},
\end{gather}
where $G_{0\BM k}^R$ is the noninteracting retarded Green's function, 
and $G_{\BM k}^R$, $G_{\BM k}^A$, and $G_{\BM k}^<$ are the full retarded, advanced, and lesser Green's functions, respectively. 
Each Green's function takes the form of the matrix in terms of the Floquet and Nambu spaces,
and ``$*$" is the matrix multiplication in their product space.
% The superconducting gap function 
The self-consistent equation for the order parameter is given by 
\begin{equation}
  \label{eq.DeltaGless}
\Delta(t)=i \frac{U}{2N} \sum_{\BM k} {\rm Tr} \left( \tau_1 G^{<}_{\BM k}(t,t) \right).
\end{equation}
Here, $\tau_1$ is the Pauli matrix acting on the Nambu space. In this way, we can take account of arbitrary higher-order harmonics of the gap oscillations self-consistently.
In practice, we put a cutoff for the Floquet matrix size, for which we have confirmed the convergence of the results numerically. 
The details of the formalism are explained in Supplemental Material (SM)~\cite{footnote1}.

\textit{Results.}---
In Fig.~\ref{Fig.1}(b), we plot the time-averaged superconducting order parameter $\Delta_0$ rescaled by the equilibrium one $\Delta_{\rm eq}$ (shown by the color code)  for different amplitudes and frequencies of the driving field.
We first comment on the results in the high-frequency regime briefly.
When the frequency $\Omega$ is greater than the other energy scales of the system,
the steady state is well described by the effective Hamiltonian in the high-frequency expansion \cite{bukov_universal_2015,Eckardt_2015,Mikami2016,oka_floquet_2019}, the leading term of which is given by the time-averaged one, 
$H^{\rm eff} = \sum_{\BM{k},\sigma}J_0(A)\varepsilon_{\BM{k}}c^{\dagger}_{\BM{k}\sigma}c_{\BM{k}\sigma}$$-\frac{U}{N}\sum_{\BM{k},\BM{p}} c^{\dagger}_{\BM{k}\uparrow}c^{\dagger}_{-\BM{k}\downarrow} c_{-\BM{p}\downarrow}c_{-\BM{p}\uparrow}$$+O(1/\Omega)$.
Here the kinetic term is renormalized by the zeroth-order Bessel function $J_0(A)$.
Since $|J_0(A)|<1$ for any nonzero amplitudes $A$, this renormalization factor suppresses the hopping amplitude and the bandwidth,
leading to dynamical localization when $J_0(A)=0$ \cite{dunlap_dynamic_1986}.
In superconductors, this suppression of the hopping relatively enhances the pairing interaction (with the effective interaction $U_{\rm eff}=U/|J(A)|$), thus enhancing superconductivity \cite{ido_correlation-induced_2017, Dasari2018} (see the upper red area of Fig.~\ref{Fig.1}(b)).
For numerical details, we refer to SM \cite{footnote1}.

In the low-frequency ($\Omega<\Delta_{\rm eq}$) and weak-field regime, we can observe another enhanced area in the lower left part of Fig.\ref{Fig.1}(b).
Such an enhancement of superconductivity with the low-frequency drive is reminiscent of the Eliashberg effect
\cite{wyatt_microwave-enhanced_1966,ivlev_nonequilibrium_1973,dayem_behavior_1967,eliashberg_film_1970,ivlev_influence_1971, Demser2020},
most of the previous arguments of which have been based on the gap equation,
\begin{equation}
    \int \frac{d\omega}{\omega}A(\omega)(1-2f(\omega))= 
    {\frac{1}{U}},
    \label{Eq.Eliashberg}
\end{equation}
with the spectral function $A(\omega)$ and distribution $f(\omega)$.
In the conventional argument, one assumes that the effect of the external field only appears as a perturbation to the Fermi-Dirac distribution, 
$f(\omega)=f_{T}^{\rm FD}(\omega)+\delta f(\omega)$,
with the spectral function being fixed to the equilibrium one ($A(\omega)=A_{\rm eq}(\omega)$), and that the gap equation (\ref{Eq.Eliashberg}) remains to be valid.
The perturbation $\delta f(\omega)$ can be separately evaluated by the quantum Boltzmann equation \cite{chang1977kinetic,de2014evidence,curtis_cavity_2019}.
Due to the weight $1/\omega$ in Eq.~(\ref{Eq.Eliashberg}), 
$\delta f(\omega)$ that is negative at lower $\omega (>0)$ and positive at higher $\omega$ is favored for enhancing superconductivity.
One can roughly attribute this effect to a shift of an effective temperature defined by $ \frac{\delta T}{T}\equiv\int_{-\infty}^{+\infty} \frac{\delta f(\omega)}{\omega} \,d\omega$.
Enhancement of superconductivity may then be expected if $\delta T$ becomes negative (i.e., effective cooling).

\begin{figure}[t]
    \centering
        \includegraphics[width=1.0\linewidth]{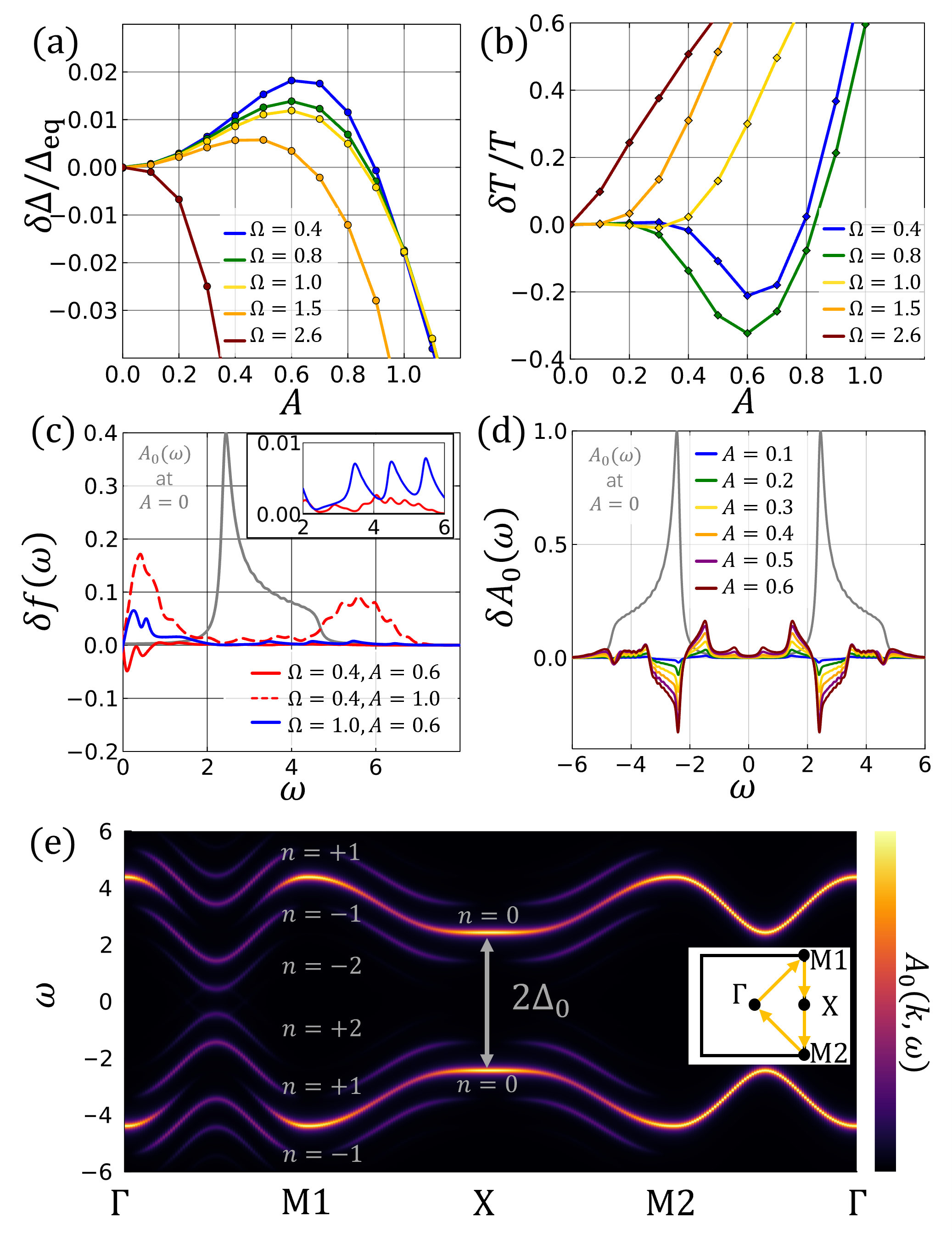} 
    \caption{
    Relative changes of (a) the superconducting order parameter $\delta\Delta/\Delta_{\rm eq}(\approx 2.4)$ and (b) the effective temperature $\delta T/T$ as a function of the driving amplitude $A$.
    (c) Changes of the effective distribution $\delta f(\omega)$.
    The inset shows an enlarged view of $\delta f(\omega)$.
    The gray solid line shows the time-averaged spectral function $A_0(\omega)$ for $A=0$, whose maximum is normalized to 0.4.
    (d) Changes of the spectral function $\delta A_0(\omega)$, which has been normalized to set the maximum of $A_0(\omega)$ at $A=0$ to $1.0$.
    (e) Momentum-dependent spectral function $A_0(\BM k,\omega)$ for $\Omega=1.0$ and $A=0.6$.
    The number $n$ marks the Floquet sidebands.
    The inset shows the path in the Brillouin zone.
    The other parameters are $U=6.0$, $\Gamma=0.1$, and $T=0.2$.
    }
    \label{Fig.3}
\end{figure}

In our formalism, Eq.~\eqref{Eq.Eliashberg} is not
strictly valid due to the strong driving and the coupling to the bath.
Instead, we compute the nonequilibrium distribution, the superconducting order parameter, and the spectral function
fully self-consistently through Eqs.~\eqref{Eq.Dyson1}-\eqref{eq.DeltaGless}.
We define the distribution change as $\delta f(\omega)\equiv A_0(\omega)/n_0(\omega) - f^{\rm FD}_{T}(\omega)$,
where $A_0(\omega)$ and $n_0(\omega)$ are the time-averaged spectral function and occupation function given by the Floquet retarded and lesser Green's functions, respectively \cite{footnote1}.

In Fig.~\ref{Fig.3}(a)-(b), we present the relative change in the time-averaged order parameter $\delta\Delta$ and effective temperature $\delta T$ across different driving frequencies. 
The results demonstrate that superconductivity is indeed enhanced for weak $A$ and $\Omega < \Delta_{\rm eq}(\approx 2.4)$, 
and such enhancement smoothly turns into suppression as $A$ increases.
Figure~\ref{Fig.3}(c) provides a detailed view of $\delta f(\omega)$.
Specifically, for $\Omega = 0.4$ and $A = 0.6$, where superconductivity is most strongly enhanced at that frequency, 
$\delta f(\omega)$ is indeed negative around $\omega = 0$, 
resulting in negative $\delta T$ in Fig.~\ref{Fig.3}(b).
In contrast, at $\Omega = 0.4$ and $A = 1.0$, where suppression of superconductivity happens, 
$\delta f(\omega)$ becomes entirely positive at $\omega>0$,
thus resulting in positive $\delta T$.
Although this distribution indicates effective heating, 
it remains to be of the Eliashberg type in the sense that quasiparticles are accumulation at the far edges of the bands (Fig.~\ref{Fig.3}(c)).

A discrepancy between the sign of $\delta\Delta$ and $\delta T$ arises for $\Omega = 1.0$ and $A = 0.6$,
where $\delta f(\omega)$ at $\omega>0$ and consequently $\delta T$ are both positive,
despite the strongest enhancement of superconductivity occurs.
In particular, we do not observe significant negative values of $\delta f(\omega)$ around the position of the coherence peak $\omega \approx \Delta_0$
(see the inset of Fig.~\ref{Fig.3}(c)),
which is in stark contrast to those obtained from the kinetic equations \cite{de2014evidence,curtis_cavity_2019}.
This indicates that enhancement of superconductivity can happen even without effective cooling of quasiparticles, and suggests another mechanism for such enhancement.

We attribute this mechanism to the modulation of the spectral function due to light driving.
%Here we  provide a qualitative interpretation based on Eq.~\eqref{Eq.Eliashberg}.
%Specifically, we introduce a perturbation 
Let us denote the change of the spectral function by $\delta A_0(\omega)$,
which becomes positive around $\omega\sim\Delta_0-\Omega$, and negative around $\omega\sim\Delta_0$ (Fig.~\ref{Fig.3}(d)).
These structures roughly correspond to the Floquet sidebands, making replicas of the coherence peak, which can be clearly observed in the momentum-resolved spectral function (Fig.~\ref{Fig.3}(e)).
If we assume that the distribution remains to be unchanged from $f^{T}_{\rm FD}(\omega)$, then
$\delta A_0(\omega)$ multiplied by a positive factor $\frac{1-2f^{T}_{\rm FD}(\omega)}{\omega}$ peaked at $\omega=0$ will give a positive contribution to the left-hand side of Eq.~\eqref{Eq.Eliashberg}.
In order to keep the left-hand side of Eq.~\eqref{Eq.Eliashberg} constant, the coherence peak should be shifted to larger $\omega$, thus enhancing the gap $\Delta_0$.
In actual situations, both the distribution and spectral functions are modulated by periodic driving, and the competition between the two effects will determine the overall change of the superconducting order parameter.

\begin{figure}[t]
    \centering
        \includegraphics[width=1.0\linewidth]{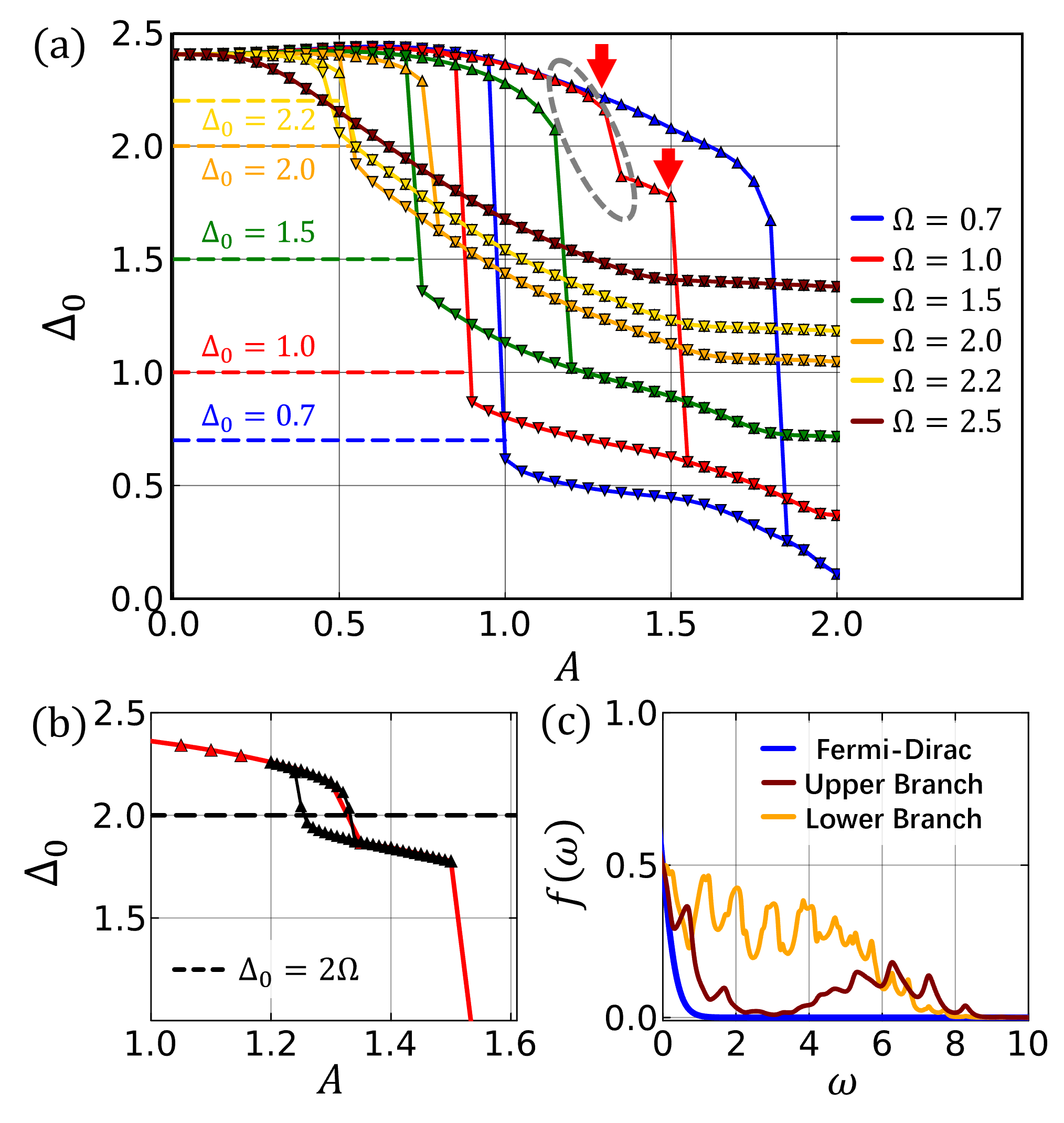} 
    \caption{
    (a) Hysteresis behavior of $\Delta_0$ as a function of $A$ with different driving frequencies. The upper (lower) branches correspond to $\Delta_0$ computed from those with smaller (larger) $A$. The horizontal dashed lines indicate $\Delta_0=\Omega$. The red arrows show sudden jumps of $\Delta_0$ for $\Omega=1.0$.
    (b) Enlarged view of $\Delta_0$ in the region marked by the gray dashed curve in (a). The black curves with triangle markers show the nested hysteresis behavior of $\Delta_0$ for $\Omega=1.0$. The black dashed horizontal line shows $\Delta_0=2\Omega=2.0$.
    (c) Effective distributions for $\Omega=1.0$ and $A=1.2$ corresponding to the upper (dark red) and lower (orange) branches of the hysteresis loop. 
    The blue curve is the Fermi-Dirac distribution with $T=0.2$ for reference.
    The other parameters are $U=6.0$, $\Gamma=0.1$, and $T=0.2$, which gives $\Delta_{\rm eq}\approx 2.4$.
    }
    \label{Fig.4}
\end{figure}

So far, we have demonstrated how the Eliashberg-type enhancement of superconductivity breaks down as the driving strength increases.
One might expect the superconducting order to diminish down to zero smoothly.
However, 
%as we show in the following,
%our further simulations reveal that 
this transition is not smooth:
In Fig.~\ref{Fig.4}(a), we plot $\Delta_0$ as a function of $A$ for different light frequencies $\Omega$. As one can see, there are two branches in the $\Delta_0$-$A$ curves: The upper branches marked with upper triangles are computed from small to large $A$,
i.e., the initial value of $\Delta_0$ for the self-consistency loop is taken to be the converged value of $\Delta_0$ at smaller $A$.
The lower branches marked with lower triangles are computed in the opposite direction.
This discontinuous transition is clearly a non-perturbative effect with respect to $A$, which cannot be captured by a naive perturbation theory.

%In Fig.\ref{Fig.4}(a), we plot the $\Delta_0$-$A$ curves with different light frequencies $\Omega$.
%The $\Delta_0$-$A$ curves show two branches.
%Technically, the upper branches marked with upper triangles are computed from smaller amplitudes to larger ones,
%i.e. the initial value of $\Delta_0$ to start the self-consistent loop is adopted from the converged value of the left neighboring point,
%and the lower branches marked with lower triangles are computed in the opposite direction.

When $\Omega$ is smaller than $\Delta_{\rm eq}$ ($\approx 2.4$),
%for the parameters used in Fig.~\ref{Fig.4}(a)), 
%the time-averaged superconducting order parameter 
$\Delta_0$
in the upper branches shows a drastic decline at certain driving amplitudes after the Eliashberg-type enhancement, and merges into the lower branches.
The lower branches show similar sudden jumps at different driving amplitudes.
For a fixed $\Omega$, the two branches form a hysteresis loop of the superconducting order parameter, indicating a first-order-like phase transition.
The hysteresis regions shrink with increasing light frequencies, and completely disappear when the frequency becomes larger than $\Delta_{\rm eq}$ (see the curves of $\Omega=2.5$ in Fig.~\ref{Fig.4}(a)).
The disappearance of the hysteresis loops at $\Omega\simeq\Delta_{\rm eq}$ suggests that the two-photon-assisted tunneling triggers the transition.
With careful observation, 
one can see in Fig.~\ref{Fig.4}(a) that for each frequency the drastic decline of the two branches happens right before cutting the horizontal dashed line of $\Delta_0=\Omega$,
which meets the two-photon absorption
condition since $2\Delta_0$ is the gap size of the superconducting NESS.

One interesting observation here is that such a first-order phase transition can be nested and triggered by higher-order processes, implying a hierarchical structure of nonequilibrium phases with multi-stability.
For instance, the upper branch of the loop with $\Omega=1.0$ in  Fig.~\ref{Fig.4}(a) has another steep decline around $A=1.3$, marked with the red arrow.
If we sweep $A$ from smaller to larger values and vice versa within the upper branch around $A=1.3$, we find a nested hysteresis loop (Fig.~\ref{Fig.4}(b)).
%We again performed the two-direction computation around $A=1.3$ and found a nested hysteresis loop, as shown in Fig.\ref{Fig.4}(b).
This secondary hysteresis loop is intersected by the horizontal line of $\Delta_0=2.0$ (Fig.~\ref{Fig.4}(b)), 
indicating that the transition is induced by a four-photon-assisted excitation process that requires $2\Delta_0=4\Omega$.

To understand the nature of each state in the two main branches in the hysteresis loop,
we plot the corresponding NESS distribution functions for $A=1.0$ and $\Omega=1.0$ in Fig.~\ref{Fig.4}(c).
The distribution of the upper branch is essentially of the Eliashberg-type, 
where excited quasiparticles are accumulated at the far band edges, showing a significant population inversion within the bands,
while the distribution of the lower branch shows the breakdown of such population inversion 
accompanied by
the massive excitation of quasiparticles.
This type of a sudden collapse of the distribution is similar to the recently discussed quantum avalanche effect \cite{han_correlated_2023} 
assisted by multi-photon processes through in-gap states in the context of dielectric breakdown of correlated insulators.

In the phase diagram (Fig.~\ref{Fig.1}(b)), the hysteresis region appears in the area surrounded by the orange dashed line with the endpoint at $A\approx 0.3$ and $\Omega\approx \Delta_{\rm eq}$.
One can smoothly connect the small $A$ and large $A$ phases by going around the endpoint on a higher frequency side.
This connectivity may come from the dissipative bath, which smoothens the coherence peaks and leaves tiny but nonzero density of states within the superconducting gap.
Additionally, the phase diagram does not show any trace of single-photon absorptions.
This is because we are focusing on the clean limit of the superconductors.
We expect the effect of impurities will further enrich the phase diagram in the frequency regime $\Omega\approx 2\Delta_{\rm eq}$, which we leave as a future problem.

\textit{Conclusion.}---
In this Letter, we have studied the Floquet steady states of superconductors coupled to heat baths under the persistent and periodic light driving
by the Floquet Green's function method.
When the driving frequency $\Omega$ is smaller than $\Delta_{\rm eq}$, we find that enhancement of superconductivity can indeed occur even without effective cooling,
which is attributed to the modulation of the spectrum in the Floquet states.
We further demonstrated nonequilibrium superconducting phase transitions accompanied by multiple (nested) hysteresis behaviors, which are induced by multi-photon-assisted tunnelings.
This process tells us how the Eliashberg-type enhancement of superconductivity breaks down in the strong driving field regime.
% \nt{[Add discussion on possible experimental realization]}

We conclude by proposing several potential experimental approaches.
In solid-state systems, one can directly observe the modifications of the energy spectrum in Floquet states of superconductors by the time-resolved ARPES with terahertz pump pulses \cite{Reimann2018, Ito2023, Boschini2024}, which would be limited in terms of the energy resolution and pump intensities but will be further developed in the near future. One can also use the transient absorption, optical nonlinearity, and tunneling spectroscopies with continuous microwave irradiation to probe the Floquet steady states and the hysteresis behaviors.
Another platform is cold-atom systems, where gauge fields, dissipation, and fermionic condensates can be effectively simulated \cite{regal_observation_2004, Diehl2008, senaratne_quantum_2018, Eckardt2017, Weitenberg2021}.

\textit{Acknowledgement.}---
The authors acknowledge T. Oka, K. Kobayashi, P. Werner, K. Ishizaka and H. Li for fruitful discussions.
This work is supported by JST FOREST (Grant No.~JPMJFR2131), JST PRESTO (Grant No.~JPMJPR2256) and JSPS KAKENHI (Grant Nos.~JP22K20350, JP23K17664, and JP24H00191).
H.Z. is supported by the Forefront Physics and Mathematics Program to Drive Transformation (FoPM), 
a World-leading Innovative Graduate Study (WINGS) Program, at the University of Tokyo.

% \appendix
% \section{If there is any}

% The \nocite command causes all entries in a bibliography to be printed out,
% whether or not they are actually referenced in the text. This is an appropriate
% for the sample file to show the different styles of references, but the authors
% most likely will not want to use it.
% \nocite{*}

% \bibliographystyle{apsrev4-2}
\bibliography{reference}% Produces the bibliography via BibTeX.

\end{document}

\preprint{APS/123-QED}

\title[Manuscript]{Supplementary materials for ``Nonequilibrium hysteretic phase transitions in periodically light-driven superconductors"}
% Force line breaks with \\

\author{Huanyu Zhang}
\email{zhang@dyn.phys.s.u-tokyo.ac.jp}
\affiliation{Department of Physics, The University of Tokyo, Hongo, Tokyo 113-0033, Japan}
\author{Kazuaki Takasan}
\email{kazuaki.takasan@phys.s.u-tokyo.ac.jp}
\affiliation{Department of Physics, The University of Tokyo, Hongo, Tokyo 113-0033, Japan}
\author{Naoto Tsuji}
\email{tsuji@phys.s.u-tokyo.ac.jp}
\affiliation{Department of Physics, The University of Tokyo, Hongo, Tokyo 113-0033, Japan}
\affiliation{RIKEN Center for Emergent Matter Science (CEMS), Wako 351-0198, Japan}

%\author{\textcolor{red}{(Please help with the correct affiliation format)}}
%\affiliation{Department of Physics, The University of Tokyo.}

%\noaffiliation

\date{\today}% It is always \today, today,
             %  but any date may be explicitly specified

%\keywords{Suggested keywords}%Use showkeys class option if keyword
                              %display desired
 \maketitle
%\tableofcontents

\section{Floquet many-body theory for BCS superconductors}
\label{Sec.FMBT}

\subsection{Keldysh Formalism}
\label{Sec.Keldysh}

We give the definition of the lesser Green's function, the retarded Green's function, and the advanced Green's function
in the real-time domain:
\begin{align}
G^{<}_{\BM{k},\alpha\beta}(t,t') & =+i \left<  \psi_{\BM{k}\beta}^\dagger(t') \psi_{\BM{k}\alpha}(t) \right>, \\
G^{R}_{\BM{k},\alpha\beta}(t,t') & =-i \theta(t-t')\left<  \{\psi_{\BM{k}\alpha}(t),\psi_{\BM{k}\beta}^\dagger(t')\} \right>,\\
G^{A}_{\BM{k},\alpha\beta}(t,t') & =+i \theta(t'-t)\left<  \{\psi_{\BM{k}\alpha}(t),\psi_{\BM{k}\beta}^\dagger(t')\} \right>.
\end{align}
$\psi^{\dagger}_{\BM{k}\alpha}$ and $\psi^{\dagger}_{\BM{k}\alpha}$ denotes the fermionic creation and annihilation operator with momentum $\BM{k}$ and internal degree of freedom $\alpha$.
In the context of superconductors, the fermionic operators $\psi$ and $\psi^{\dagger}$
stand for the Nambu spinor.
$\langle ...\rangle$ stands for taking the average with respect to the initial density matrix $\rho_0$.
$\theta(t-t')$ is the step function, and the $\{...,...\}$ is the anti-commutator.

The retarded Green's function includes information on the energy spectrum (real part) and the lifetime (imaginary part) of the quasiparticles,
while the lesser Green's function contains information on the occupation of quasiparticles.
In equilibrium systems, they are connected through the Fluctuation-Dissipation theorem (DFT) \cite{haug2008quantum}, which says:
\begin{equation}
G_{\BM k}^{<}(\omega)=-2i\cdot f^{\rm FD}_{T}(\omega)\cdot {\rm Im} G_{\BM k}^{R}(\omega),
\label{DFToriginal}
\end{equation}
where $\omega$ comes from the Fourier transformation on $(t-t')$ and $f^{\rm FD}_{T}(\omega)=\frac{1}{e^{\omega}/k_BT + 1}$ is the Fermi-Dirac distribution at temperature $T$.
In non-equilibrium interacting systems, the relation above is usually invalid, and we have to solve $G^{<}_{\BM{k},\alpha\beta}(t,t')$ and $G^{R}_{\BM{k},\alpha\beta}(t,t')$ respectively.

Denoting non-interacting and interacting Green's functions with $G^{<}_{0\BM{k},\alpha\beta}(t,t')$ and $G^{R}_{0\BM{k},\alpha\beta}(t,t')$ respectively, we have the Dyson-Keldysh equations \cite{jauho_time-dependent_1994} for non-equilibrium interacting systems:
\begin{gather}
    \label{Eq.Dyson1}
    G^{R}_{\BM{k}}=G^{R}_{0\BM{k}}+G^{R}_{0\BM{k}}*\Sigma^{R}_{\BM{k}}*G^{R}_{\BM{k}},\\
    \label{Eq.Kel1}
    G^{<}_{\BM{k}}=G^{R}_{\BM{k}}*\Sigma^{<}_{\BM{k}}*G^{A}_{\BM{k}},
\end{gather}
where the multiplication stands for the abbreviation of time convolution and matrix index contraction.
For example,
\begin{equation}
\label{eq.multirule}
    (A * B)_{\alpha\beta}(t,t')\equiv \sum_{\gamma}\int {\rm d}\bar{t} A_{\alpha\gamma}(t,\bar{t})B_{\gamma\beta}(\bar{t},t').
\end{equation}

For the BCS model used in the main text,
\begin{equation}
\begin{aligned}
  H_{\rm{BCS}}(t)=\sum_{\BM{k},\sigma} & \varepsilon_{\BM{k}-e\BM{A}(t)} c_{\BM{k}\sigma}^\dagger c_{\BM{k}\sigma}\\
   &-\frac{U}{N} \sum_{\BM{k},\BM{p}} c_{\BM{k}\uparrow}^\dagger c_{-\BM{k}\downarrow}^\dagger  c_{-\BM{p}\downarrow} c_{\BM{p}\uparrow},
   \end{aligned}
\end{equation}
we adopt the standard mean-field approximation by introducing the superconducting order parameter
\begin{equation}
\begin{aligned}
\label{Eq.OrderParameter}
\Delta(t)\equiv -\frac{U}{N}\sum_{\BM{k}} & \langle c_{-\BM{k}\downarrow}(t)  c_{\BM{k}\uparrow}(t) \rangle \\
&=i \frac{U}{2N} \sum_{\BM k} {\rm Tr} \left( \tau_1 G^{<}_{\BM k}(t,t) \right).
\end{aligned}
\end{equation}
The assumption that $\Delta(t)$ is real has been included in the second equation.
The mean-field Hamiltonian can then be reshaped in a matrix form with the Nampu spinor:
\begin{align}
  \label{eq.MFHamil}
  H_{\rm{MF}}(t) =\sum_{\BM k}\psi_{\BM k}^\dagger H_{0\BM k}(t) \psi_{\BM k} +\frac{N}{U}\Delta^\dagger(t)\Delta(t),\\
  \label{eq.Hamil}
  H_{0\BM k}(t) =\left(
  \begin{array}{cccc}
  \varepsilon_{\BM{k}-e\BM{A}(t)}& \Delta(t)\\
  \Delta(t)^{\dagger} & -\varepsilon_{-\BM{k}-e\BM{A}(t)}\\
  \end{array}
  \right).
\end{align}
In the rest of the text, we drop the unimportant second term in Eq.\eqref{eq.MFHamil}.
The mean-field Hamiltonian is then formally non-interacting,
and the corresponding Green's function is denoted by $G^{R/<}_{0\BM{k}}$.

\subsection{Heat Bath}
\label{ModelBath}

In the main text, we adopted a free fermion bath model for each site of the superconductor,
whose Hamiltonian is simply
\begin{equation}
    {H}_{\rm bath}=\sum_{i,\delta} \phi_{i\delta}^{\dagger} \varepsilon^{\rm bath}_{i\delta}\phi_{i\delta}.
\end{equation}
$\phi_{i\delta}$ is the fermionic operator of the bath at the lattice site $i$ and $\delta$ denotes the internal degree of freedom.
We then assume a B{\"u}tticker's model, where the bath couples to the electrons through 
\begin{equation}
    {H}_{\rm coup}=\sum_{i\delta\alpha}\left(\psi_{i\alpha}^{\dagger}(V_{i})_{\alpha\delta}\phi_{i\delta}+\phi_{i\delta}^{\dagger}(V^{\dagger}_{i})_{\delta\alpha} \psi_{i\alpha}\right).
\end{equation}
By analytically integrating out the bath degrees of freedom,
coupling to the baths can be addressed with self-energy
\begin{equation}
    \Sigma_{i,\alpha\beta}^{R/<}=\sum_{\delta\gamma}\left[(V_i)_{\alpha\delta}\cdot G^{{\rm bath},R/<}_{i,\delta\gamma}\cdot (V^{\dagger}_i)_{\gamma\beta}\right],
\end{equation}
where $G^{{\rm bath},R/<}_{i,\delta\gamma}$ is the Green's function of the heat bath.
The baths have a considerably large number of degrees of freedom so they can be treated as is in equilibrium and have a definite temperature $T$.
% For a non-interacting fermion bath, its Green's function reads
% \begin{equation}
%     G^{{\rm bath},R}_{\BM{k},\delta\gamma}(\omega)=\frac{\delta_{\delta\gamma}}{\omega+\mu^{\rm bath}-\varepsilon_{\BM{k}\delta}^{\rm bath}+i\eta},
% \end{equation}
% in the frequency regime.
As is assumed in many previous studies \cite{tsuji_nonequilibrium_2009,aoki_nonequilibrium_2014},
the chemical potential of the bath $\mu^{\rm bath}$ can be well adjusted so that ${\rm Re} \,G^{{\rm bath},R}_{\BM{k},\delta\gamma}(\omega)=0$, which ensures no net particle flow between the bath and the superconductors.
The bath is further assumed to have a constant density of states and doesn't mix up the degrees of freedom of the Nambu spinors.
Therefore, we can neglect the frequency dependence of the imaginary part of the self-energy function and denote it with a coupling strength constant $\Gamma$, that is
\begin{equation}
    \Sigma_{\BM{k}}^{R}(\omega)=-i\Gamma\cdot \BM{1},
\end{equation}
the lesser component of the self-energy function is then given according to the fluctuation-dissipation theorem
\begin{equation}
    \Sigma_{\BM{k}}^{<}(\omega)=2if^{\rm FD}_{T}(\omega)\Gamma \cdot \BM{1}.
\end{equation}
Note that $\BM{1}$ is the $2\times2$ unit matrix in the Nambu space.

\subsection{Floquet Green's function}
\label{SecFloquet}

For a periodic non-equilibrium many-body system,
its discrete time-translation symmetry ensures the Green's functions satisfy $G^{R/<}_{\BM{k}}(t,t')=G^{R/<}_{\BM{k}}(t+\mathcal{T},t'+\mathcal{T})$, where $\mathcal{T}$ is the periodic.
In terms of $t_{\rm dif}\equiv t-t'$ and $t_{\rm ave}\equiv(t+t')/2$,
we may perform the following transformation to derive the Floquet Green's function \cite{tsuji_correlated_2008}:
% \begin{widetext}
\begin{equation}
\begin{aligned}
\label{Eq.FloquetTransformation}
 G^{R/<}_{\BM{k},mn}(\omega)\equiv  &\frac{1}{\mathcal{T}} \int^{\mathcal{T}}_{0} {\rm d}t_{\rm ave} e^{i(m-n)\Omega t_{\rm ave}}\\
&\int_{-\infty}^{\infty} {\rm d}t_{\rm dif} e^{i(\omega+(m+n)\Omega/2) t_{\rm dif}} G^{R/<}_{\BM k}(t,t'),
\end{aligned}
\end{equation}
% \end{widetext}
where $m,n$ are the Floquet indices with integer values and $\omega$ is the frequency (energy) that takes value from the Floquet Brillouin Zone $(-\Omega/2,\Omega/2]$ (add references).
The transformation above is essentially a continuous Fourier transformation for $t_{\rm dif}$ and a discrete one for $t_{\rm ave}$.
In this sense, the transformation Eq.\eqref{Eq.FloquetTransformation} keeps the multiplication structure in the real-time domain,
while the multiplication rule now reads \cite{tsuji_correlated_2008}:
\begin{equation}
\label{eq.multiruleFloquet}
    (A * B)_{\alpha\beta,mn}(\omega)\equiv \sum_{\gamma,p} A_{\alpha\gamma,mp}(\omega)B_{\gamma\beta,pn}(\omega),
\end{equation}
which is formally the multiplication rule for the direct product matrices with Nambu indices $\alpha,\beta,\gamma$ and the Floquet indices $m,n,p$.

In the matrix form, computing the inverse of Green's function becomes straightforward.
The equation of motion of the non-interacting retarded Green's function $G^{R}_{0\BM{k}}(t,t')$, which reads
\begin{equation}
\label{Eq.EOM}
    (i\partial_t - H_{0\BM{k}}(t))G^{R}_{0\BM{k}}(t,t')=\delta(t-t')\cdot \BM{1},
\end{equation}
can be addressed in terms of the Floquet Green's function,
which reads
\begin{equation}
\label{Eq.FloquetEOM}
\sum_{p}\left[(\omega+m\Omega)\delta_{mp}-H_{0\BM{k},mp}\right]G^{R}_{0\BM{k},pn}(\omega)=\delta_{mn}\cdot \BM{1},
\end{equation}
by performing the transformation Eq.\eqref{Eq.FloquetTransformation} on both sides of Eq.\eqref{Eq.EOM}.
Note that $H_{0\BM{k}}(t)G^{R}_{0\BM{k}}(t,t')= H_{0\BM{k}}(t)\delta(t-\bar{t}) * G^{R}_{0\BM{k}}(\bar{t},t')$,
so
$H_{0\BM{k},mn}= (1/\mathcal{T})\int^{\mathcal{T}/2}_{\mathcal{T}/2} {\rm d}t e^{i(m-n)\Omega t} H_{0\BM{k}}(t)$,
where $\mathcal{T}=2\pi/\Omega$ is the period.
Then, the inverse of the retarded Floquet Green's function reads
\begin{equation}
\label{eq.GRinv}
    (G^{R-1}_{0\BM{k}})_{mn}(\omega)=(\omega+m\Omega+i\eta)\delta_{mn} \cdot \BM{1}-H_{0\BM{k},mn},
\end{equation}
where the infinitesimal $\eta$ originates from the causal nature of the retarded Green's function.
It is then convenient to express the Dyson equation with the inverse of the Floquet Green's functions, that is
\begin{equation}
\label{eq.Dyson2}
    G_{\BM{k}}^{R-1}=G_{0\BM{k}}^{R-1}-\Sigma_{\BM{k}}^{R}.
\end{equation}
Note that the self-energy of the bath with Floquet indices reads:
\begin{gather}
     \Sigma_{\BM{k},mn}^{R}(\omega)=-i\Gamma\cdot \delta_{mn}\BM{1},\\
     \label{Eq.BathFloquet2}
     \Sigma_{\BM{k},mn}^{<}(\omega)=2if^{\rm FD}_{T}(\omega+m\Omega)\Gamma \cdot \delta_{mn} \BM{1},
\end{gather}
where $\omega$ lies in the Floquet BZ: $(-\Omega/2,\Omega/2]$.

In numerical computations, we truncate the Floquet indices in numerical calculation with $|m|,|n|<N$, 
which allows us to solve the Dyson-Keldysh equations with finite-sized matrix inverse and multiplication. 
The error from the truncation is analytically well-bounded \cite{rudner_floquet_2020}.
In this study, the truncation is practically always set to meet the condition of $N\Omega>W$, where $W$ is the bandwidth read from the spectral function, to minimize the error.
The truncation $N$ becomes very large in the low-frequency regime, which is the main difficulty in the numerical computation.

We comment that no coupling to the baths ($\Gamma=0$) seemingly results in an illness of the Eq.\eqref{Eq.Kel1},
that is because a finite $\Gamma$ wipes out the information of the initial state.
Such illness can be effectively cured by introducing $\Sigma_{\BM{k},mn}^{<}(\omega)=2if(\omega+m\Omega)\delta_{mn}\cdot\eta$,
where $\eta \rightarrow 0+$ and $f(\omega)$ is the distribution of the initial state that can be nonthermal in principle \cite{Eckstein2018}.
However, it is well known that such a isolated system will eventually be heated to a featureless state, 
which can be technically reflected by the increasing numerical instability when $\Gamma$ approaches zero.

\subsection{Self-consistent Equation for the Order Parameter}
\label{algorithm}

The order parameter needs to be self-consistently determined by Eq.\eqref{Eq.OrderParameter}.
We express Eq.\eqref{Eq.OrderParameter} in terms of the Floquet Green's function. 
Noting that
\begin{equation}
\begin{aligned}
G^{<}_{\bm{k}}(t,t)&=\lim_{\delta t \rightarrow 0} G^{<}_{\bm{k}}(t,t-\delta t)\\
% =& \lim_{\delta t \rightarrow 0} \frac{1}{2\pi}\sum_{mn}\int^{\Omega/2}_{-\Omega/2}{\rm d}\omega e^{-i(\omega+m\Omega)t+i(\omega+n\Omega)(t-\delta t)} G^{<}_{\bm{k},mn}(\omega)\\
&=\frac{1}{2\pi}\sum_{mn} e^{-(m-n)\Omega t} \int^{\Omega/2}_{-\Omega/2}{\rm d}\omega G^{<}_{\bm{k},mn}(\omega),
\end{aligned}
\end{equation}
and
$\Delta_{mn}=\Delta_{m-n}\equiv (1/\mathcal{T})\int^{\mathcal{T}/2}_{\mathcal{T}/2} {\rm d}t e^{i(m-n)\Omega t} \Delta(t)$,
we have
\begin{equation}
    \label{Eq.DeltaFloquet}
    % \begin{aligned}
        \Delta_{mn}=i\frac{U}{2\pi N}\sum_{\bm k} \sum_{p,q}^{p-q=m-n} \int_{-\Omega/2}^{+\Omega/2}{\rm d}\omega G^{<}_{\bm{k},pq}(\omega).
    % \end{aligned}
\end{equation}
% \begin{equation}
%     \begin{aligned}
%         \Delta_{mn}=&\Delta_{m-n}=\frac{\Omega}{2\pi}\int_{-\pi/\Omega}^{+\pi/\Omega}{\rm d}t e^{i(m-n)t} \\ 
%        \cdot &\left[i\frac{U}{2\pi N} \sum_{\bm k,p,q}e^{-i(p-q)\Omega t}
%         \int_{-\Omega/2}^{+\Omega/2}{\rm d}\omega G^{<}_{\bm{k},pq}(\omega)\right]\\
%         =&i\frac{U}{2\pi N}\sum_{\bm k} \sum_{p,q}^{p-q=m-n} \int_{-\Omega/2}^{+\Omega/2}{\rm d}\omega G^{<}_{\bm{k},pq}(\omega).
%     \end{aligned}
%     \label{DeltaFloquet}
% \end{equation}

Now, Eq.\eqref{Eq.Kel1},\eqref{eq.Dyson2},\eqref{Eq.DeltaFloquet} form the self-consistent equations for the order parameter.
We solve them by iteration.
We start with an initial order parameter $\Delta^{\rm ini}_{mn}$,
which is assumed to be static, i.e. $\Delta^{\rm ini}_{0}\neq 0$ and $\Delta^{\rm ini}_{mn}=0$ with $m\neq n$.

Substituting $\Delta^{\rm ini}_{mn}$ into the Hamiltonian $H_{0\BM{k},mn}$ and further into Eq.\eqref{eq.GRinv},
we have the non-interacting retarded Green's function $G^{R}_{0\BM{k},mn}(\omega)$.
Taking the inverse of the right-hand side of the Dyson equation Eq.\eqref{eq.Dyson2}, we further derive the interacting retarded Green's function $ G^{R}_{\BM{k},mn}(\omega)$.
Note that the advanced Green's function satisfies $G^{A}_{\BM{k},mn}(\omega)=(G^{R}_{\BM{k}})^{\dagger}_{mn}(\omega)$,
the lesser Green's function $ G^{<}_{\BM{k},mn}(\omega)$ can then be derived from the Keldysh equation Eq.\eqref{Eq.Kel1}.
Next, we use the right-hand side of Eq.\eqref{Eq.DeltaFloquet} to update the order parameter.
The procedure above is repeated until we have a converged result for $\Delta_{mn}$.

The convergence for the superconducting states with $\Delta_{0}\neq0$ is typically quick,
while the convergence for the normal states is not. 
We practically have a maximum time of iteration for the normal state, 
which makes their $\Delta_0$ tiny but nonzero.

In the main text, we focus mainly on the zeroth order component $\Delta_{0}$, the time-averaged order parameter, as the indicator of superconductivity in the non-equilibrium steady states (NESS).
We also compute the time-averaged spectral function $A_{0}(\omega)$ and occupation function $n_{0}(\omega)$ for quasiparticles.
\begin{gather}
\label{eq.spectrum}
   A_{0}(\omega)=-\frac{1}{\pi N} \sum_{\BM{k}}{\rm Im} \left[ {\rm Tr} \left(G^{R}_{\BM{k},mm}(\omega-m\Omega)\right) \right],\\
   n_{0}(\omega)=\frac{1}{2\pi N i} \sum_{\BM{k}}{\rm Tr} \left(G^{<}_{\BM{k},mm}(\omega-m\Omega)\right).
\end{gather}
The trace in the formulas above acts on Nambu indices, 
and the Floquet index $m$ is determined by the condition that $(\omega-m\Omega)$ falls inside the Floquet BZ $(-\Omega/2,\Omega/2]$.
In the main text, we introduced the effective distribution function for the NESS:
\begin{equation}
    \label{eq.effectivedistribution}
    f^{\rm eff}(\omega)\equiv \frac{n_0(\omega)}{A_0(\omega)}.
\end{equation}
The definition is inspired by the DFT Eq.\eqref{DFToriginal}, and the effective distribution reduces to Fermi-Dirac distribution $f^{\rm FD}_{T}(\omega)$ if the system stays in equilibrium states, like the situation when the light driving is off ($A=0$).

\begin{figure}[b]
    \centering
    \begin{subfigure}[t]{0.9\linewidth} 
        \centering
        \includegraphics[width=\textwidth]{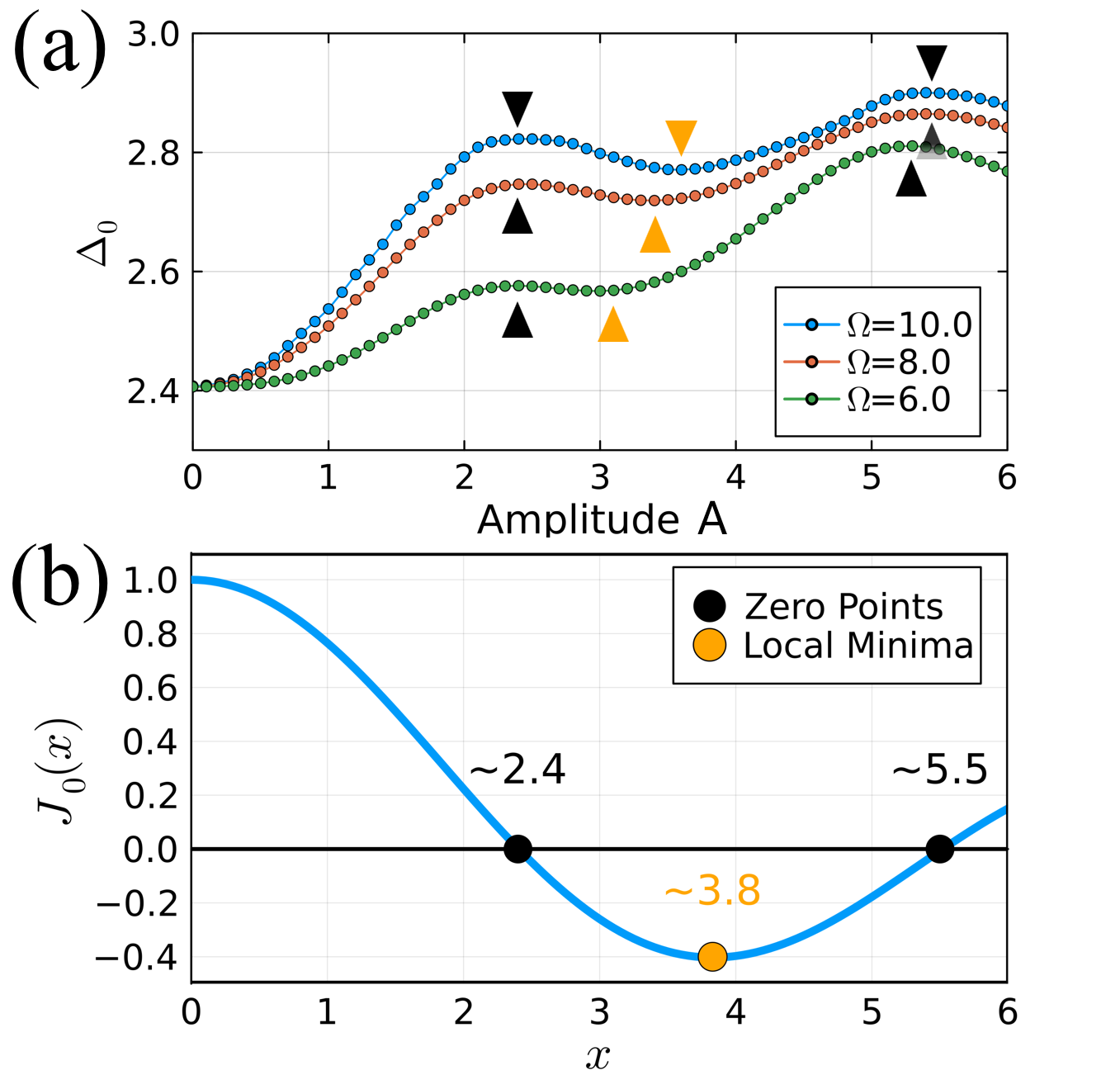} 
    \end{subfigure}
    \captionsetup{justification=RaggedRight, singlelinecheck=true, font=small}
    \caption{
    (a) The time-averaged order parameter $\Delta_0$ as a function of $A$, computed with $\Omega=6.0,8.0,10.0$ respectively.
    The black triangles mark the local maximum of $\Delta_0$ and the yellow triangles mark the local minimum of $\Delta_0$.
    The results are computed with $U = 6.0$, $\Gamma= 0.1$, $T = 0.2$ and the truncation of the Floquet indices $N=10$.
    (b) The zeroth order Bessel function $J_0(x)$. The black circles mark its two zero points at $2.4$ and $5.5$  approximately.
    The yellow circle marks its local minium at $3.8$ approximately.
    }
    \label{Fig.High}
\end{figure}

\section{Numerical Results in the High-Frequency Regime}
\label{Sec.High}

In the main text, we stated that when the frequency $\Omega$ is greater than the other energy scales of the system,
the steady state is then well described by the effective Hamiltonian in the high-frequency expansion \cite{oka_floquet_2019}, the leading term of which is given as
\begin{equation} 
\begin{aligned}
H^{\rm eff} = \sum_{\BM{k},\sigma}&J_0(A)\varepsilon_{\BM{k}}c^{\dagger}_{\BM{k}\sigma}c_{\BM{k}\sigma}\\
-&\frac{U}{N}\sum_{\BM{k},\BM{p}} c^{\dagger}_{\BM{k}\uparrow}c^{\dagger}_{-\BM{k}\downarrow} c_{-\BM{p}\downarrow}c_{-\BM{p}\uparrow}
 +O(1/\Omega).
 \end{aligned}
\end{equation}
When the frequency $\Omega$ is even larger than the band width, the lacking of scattering channels in the current model suppresses the heating inside the band,
the distribution $f^{\rm eff}(\omega)$ thus remains the Fermi-Dirac one in the range of the energy $(-\Omega,\Omega)$.

We plot the relation between $\Delta_0$ and $A$ in Fig.\ref{Fig.High}(a).
We can see the non-monotonic increase of $\Delta_0$ with respect to $A$.
The local maximum of $\Delta_0$ agrees with the zero points of $J_0(A)$ at $2.4$ and $5.5$.
With the frequency $\Omega$ increasing,
and the local minimum of $\Delta_0$ approaches the extreme points of $J_0(A)$ of $3.8$.
This shows that $\Delta_0$ follows the effective pairing strength $U^{\rm eff}=U/|J_0(A)|$ with satisfactory.

One problem of realizing such enhancement with high-frequency driving is apparently the heating from other scattering channels.
Recently, it is proposed to utilize the cavity photons instead, which can realize the similar dynamical localization as well as suppressing the heating.

% The \nocite command causes all entries in a bibliography to be printed out,
% whether or not they are actually referenced in the text. This is an appropriate
% for the sample file to show the different styles of references, but the authors
% most likely will not want to use it.
% \nocite{*}

\bibliographystyle{apsrev4-2}
\bibliography{reference}% Produces the bibliography via BibTeX.

% \twocolumn